\documentclass[cup5b]{caps}

\usepackage{graphicx}

\begin{document}

\pagenumbering{arabic}

\author[]{MASAYUKI UMEMURA\\
Center for Computational Physics, University of Tsukuba}

\chapter{ The Growth of Supermassive Black Holes and QSO Formation }

\begin{abstract}
A novel mechanism to build up a supermassive black hole (SMBH) 
is proposed.
Here, relativistic radiation-hydrodynamic processes are considered
from a galactic scale to a horizon scale.
In the present scenario, the mass of a SMBH is predicted to be in proportion
to the bulge mass, and it turns out that 
the SMBH-to-bulge mass ratio is basically determined by 
the nuclear energy conversion efficiency from hydrogen to helium, 
$\varepsilon=0.007$. The BH hierarchy is also addressed in relation to
the BH formation in globular clusters.
Futhermore, a new picture for QSO formation is proposed based on the present SMBH 
formation scenario. It is predicted that 
a host luminosity-dominant ``proto-QSO phase'' exists before the QSO phase,
and the proto-QSO phase is preceded by an optically-thick ultraluminous infrared
galaxy (ULIRG) phase.
\end{abstract}

\newcommand{\re}{\par\hangindent=0.5cm\hangafter=1\noindent}

\section{Introduction}

The recent discovery of high redshift quasars 
with $z>6$  (Fan et al. 2001) implies that
the formation of supermassive black holes proceeded in less than $10^9$ yr.
Also, the recent compilation of the kinematical
data of galactic centers in both active and inactive galaxies 
has shown that a central 'massive dark object' (MDO), 
which is the nomenclature for a black hole (BH) candidate, 
correlates with the properties of galactic bulges.
The demography of MDOs have revealed the following relations:
\re
1) The BH mass exhibits a linear relation to the bulge mass with
the ratio of 
\begin{equation}
f_{\rm BH}\equiv {M_{\rm BH} \over M_{\rm bulge}} =0.001-0.006
\end{equation}
\hspace*{4.5mm}
as a median value (Kormendy \& Richstone 1995; Richstone et al. 1998; 
Magorrian et al. 1998; Gebhardt et al. 2000a;
Ferrarese \& Merritt 2000; Merritt \& Ferrarese 2001a; McLure \& Dunlop 2002).
\re
2) The BH mass correlates with the velocity dispersion of bulge
stars with a power-law relation as $M_{\rm BH} \propto \sigma^n$,
$n=3.75$ (Gebhardt et al. 2000b), 4.72 (Ferrarese \& Merritt 2000;
Merritt \& Ferrarese 2001a, b), or $4.02 \pm 0.32$ (Tremaine et al. 2002). 
\re
3) The $f_{\rm BH}$ tends to grow with the age of youngest stars in a bulge 
until $10^9$ yr (Merrifield et al. 2000).
\re
4) In disk galaxies, the mass ratio is significantly smaller than 0.01
if the disk stars are included (Salucci 2000; Sarzi et al. 2001). 
\re
5) For quasars the $f_{\rm BH}$ is of a similar level to 
that for elliptical galaxies (Laor 1998; Shields et al. 2002).
\re
6) The $f_{\rm BH}$ in Seyfert 1 galaxies is not well converged, which may be 
considerably smaller than 0.001 (Wandel 1999; Gebhardt et al. 2000a) or
similar to that for ellipticals (McLure \& Dunlop 2002), 
while the BH mass-to-velocity dispersion relation 
in Seyfert 1 galaxies seems to hold in a similar way to elliptical 
galaxies (Gebhardt et al. 2000a; Nelson 2000).


On the other hand, the X-ray emission (Brandt et al. 1997) 
or Pa$\alpha$ lines (Veilleux, Sanders, \& Kim 1999) intrinsic for 
active nuclei have been detected in roughly one forth of ultraluminous 
infrared galaxies (ULIRGs). Furhtermore, 
it has been revealed that QSO host galaxies 
are mostly luminous and well evolved early-type galaxies
(McLeod \& Rieke 1995; Bahcall et al. 1997; Hooper, Impey, \& Foltz 1997;
McLeod, Rieke, \& Storrie-Lombardi 1999; 
Brotherton et al. 1999; Kirhakos et al. 1999;
McLure et al. 1999; McLure, Dunlop, \& Kukula 2000). 
All these findings on QSO hosts and supermassive BHs
imply that the formation of a QSO, a bulge, and a supermassive BH 
is mutually related.

Some theoretical models have been hitherto considered to explain the
BH-to-bulge correlations (Silk \& Rees 1998; 
Adams, Graff, \& Richstone 2001; Ostriker 2000).
But, little has been elucidated regarding the physics on
the angular momentum transfer which is requisite for BH formation,
since the rotation barrier by the tidal spin up 
in a growing density fluctuation is given by
\begin{equation}
{R_{\rm barr} \over R_{\rm Sch}}
\approx 10^7 \left({M_b \over 10^{8} M_\odot}\right)^{-2/3}
\left({\lambda \over 0.05}\right)^2(1+z)^{-1}
\end{equation}
in units of the Schwarzshild radius $R_{\rm Sch}$,
where $M_b$ is the baryonic mass, $z$ is the cosmological redshift,
and $\lambda$ is the spin parameter which
provides the ratio of circular velocity to velocity dispersion of dark matter
(Umemura 2001).
Furthermore, required mechanisms for BH formation must work
effectively in a spheroidal system like a bulge.
The $\alpha$-viscosity or non-axisymmetric gravitational instabilities 
would effectively transfer angular momentum 
once a disk-like system forms, but they are not likely to work
in a spheroidal system.

In this paper, as a potential mechanism that works in a spheroidal system, 
the relativistic drag force by the radiation from bulge stars is considered.
As a result, the BH-to-bulge ratio is basically determined 
by the energy conversion efficiency for nuclear fusion 
of hydrogen to helium, $\varepsilon=0.007$.
Also, in relation to BH growth, a new scenario for QSO formation 
is addressed.

\section{Formation of Supermassive Black Holes}

\subsection{Mass Accretion due to Radiation Drag}

A radiation hydrodynamic model which could account for the putative correlations 
between SMBHs and bulges is recently proposed by Umemura (2001),
where the relativistic drag force by the radiation from bulge stars 
is considered. 
The radiation drag can extract angular momentum from gas and allow the gas
to accrete onto the center 
(Umemura, Fukue, \& Mineshige 1997, 1998; 
Fukue, Umemura, \& Mineshige 1997). 
For the total luminosity $L_*$ of a uniform bulge, 
the radiation energy density is given by 
\begin{equation}
E \simeq L_*/cR^2, 
\end{equation}
where $c$ is the light speed and $R$ is the radius of the bulge.
Then, the angular momentum loss rate by the radiation drag is given by 
\begin{equation}
d \ln J / dt \simeq - \chi E / c,
\end{equation}
where $J$ is the total angular momentum of gaseous component and
$\chi$ is the mass extinction coefficient which is given by $\chi = \kappa_d /\rho$
with dust absorption coefficient $\kappa_d$ and gas density $\rho$.
Therefore, in an optically-thin regime, 
\begin{equation}
d \ln J / dt \simeq - { \tau L_* \over c^2 M_g},
\end{equation}
where $\tau$ is the total optical depth of the system
and $M_g$ is the total mass of gas.
In an optically-thick regime, the radiation drag efficiency is saturated
due to the conservation of the photon number (Tsuribe \& Umemura 1997). Thus, 
an expression of the angular momentum loss rate 
which includes both regimes is given by
\begin{equation}
d \ln J / dt \simeq - {L_* \over c^2 M_g} (1-{\rm e}^{-\tau}).
\end{equation}
Then, the mass accretion rate is estimated to be
\begin{equation}
\dot{M} = -M_g{d \ln J \over dt} \simeq {L_* \over c^2}(1-{\rm e}^{-\tau}).
\end{equation}
In an optically-thick regime, this gives simply 
\begin{equation}
\dot{M} = {L_* \over c^2},
\end{equation}
which is numerically 
$
\dot{M} = 0.1 M_\odot {\rm yr}^{-1} (L_* / 10^{12} L_\odot).
$
This rate is comparable to the Eddington mass accretion rate for
a black hole with $10^8 M_\odot$, that is,
\begin{equation}
\dot{M}_{\rm Edd} = 0.2 M_\odot {\rm yr}^{-1} \eta^{-1} 
\left(M_{\rm BH} \over 10^{8} M_\odot \right), \label{m-dot-edd}
\end{equation}
where $\eta$ is the energy conversion efficiency. Unless otherwise stated, 
$\eta=0.42$ for an extreme Kerr black hole is assumed.
The timescale of radiation drag-induced mass accretion is
\begin{equation}
t_{\rm drag} \simeq {c^2 R^2 \over \chi L_*} = 8.6 \times 10^7{\rm yr} 
R_{\rm kpc}^2 \left(L_* \over 10^{12} L_\odot \right)^{-1}
\left(Z \over Z_\odot \right)^{-1},
\end{equation}
where $R_{\rm kpc}=R/{\rm kpc}$ and $Z$ is the metallicity of gas.
It is noted that the gas which is more abundant in metals accretes 
in a shorter timescale, because the extinction is predominantly given
by the dust opacity.
For the moment, an optically-thick stage is considered.
Due to the mass accretion induced by the radiation drag,
a massive dark object (MDO) forms at the center of bulge.
Then, the mass of MDO is estimated in terms of
\begin{equation}
M_{\rm MDO} =\int^t_0 \dot{M} dt \simeq \int^t_0 
{L_* \over c^2} dt. \label{m-mdo}
\end{equation}
In practice, it is likely that optically-thin surface layers are stripped 
from optically-thick clumpy clouds by the radiation drag, and
the stripped gas losing angular momentum accretes onto the center
(Kawakatu \& Umemura 2002).

Next, we employ a simplest analytic model for bulge evolution.
The star formation rate is assumed to be 
a Schmidt law, $ S(t)=kf_g$.
If we invoke the instantaneous recycling approximation,
the star formation rate is given by
\begin{equation}
\dot{M}_*/M_b=k{\rm e}^{-\alpha kt},
\end{equation}
where $\alpha$ is the net efficiency of
the conversion into stars after subtracting the mass loss.
The radiation energy emitted by a main sequence star is
$0.14\varepsilon$ to the rest mass energy of the star, $m_* c^2$,
where $\varepsilon$ is the energy conversion efficiency of 
nuclear fusion from hydrogen to helium, which is 0.007.
Thus, the luminosity of the bulge is estimated to be
\begin{equation}
L_*=0.14\varepsilon k{\rm e}^{-\alpha kt}M_b c^2.
\end{equation}
By substituting this in (\ref{m-mdo}), 
\begin{equation}
M_{\rm MDO}=0.14\varepsilon \alpha^{-1}M_b(1-{\rm e}^{-\alpha kt}). 
\label{m-mdo2}
\end{equation}
The term $M_b(1-{\rm e}^{-\alpha kt})$ represents just the stellar 
mass in the system which is $M_{\rm bulge}$ observationally.
As a consequence, the MDO mass to bulge mass ratio is given by
\begin{equation}
{M_{\rm MDO} \over M_{\rm bulge}}=0.14\varepsilon
\alpha^{-1}=0.002\alpha_{0.5}^{-1}, 
\end{equation}
where $\alpha_{0.5}=\alpha/0.5$.
It should be noted that the final mass is basically determined by 
$\varepsilon$.
Also, the proportionality of SMBH mass to bulge mass is a natural
consequence in the present mechanism.

\begin{figure}
    \centering
  \vspace{0.5cm}
\includegraphics[width=8cm]{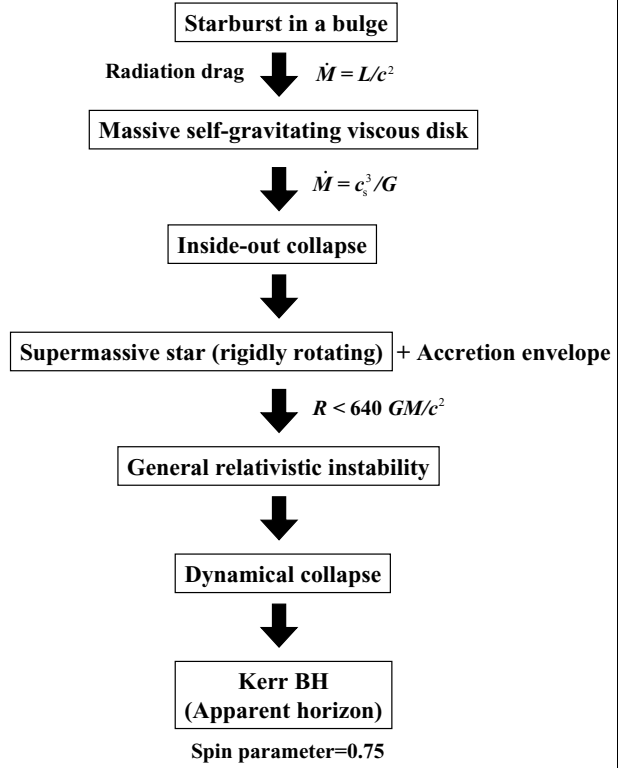}
    \caption{
A picture for the formation of a supermassive 
black hole from a galactic scale to a horizon.
The radiation from a starburst in a bulge exerts the radiation drag
onto dusty interstellar gas. Resultantly, the interstellar gas sheds
angular momentum and accrete onto the center to form a massive dark object 
(MDO). The MDO is equivalent to a massive self-gravitating viscous disk.
This disk undergoes the inside-out viscous collapse to
form a rigidly rotating supermassive star with accretion envelope.
A rigidly rotating supermassive star becomes subject to 
the general relativistic instability. Then, the supermassive
star collapse dynamically and eventually a Kerr black hole is born.}
    \label{fig1}
\end{figure}

\subsection{Towards the Horizon}

In the above, the prediction for the MDO-to-bulge mass relation is made
in the context of radiation drag-induced mass accretion.
However, the MDO itself is not a supermassive black hole, because
the radiation drag is not likely to remove the angular momentum
thoroughly, but some residual angular momentum will terminate
the radial contraction. Hence, we should consider the further
collapse of the MDO through other physical mechanisms. 

In the MDO, the viscosity is expected to work effectively because 
the timescale for viscous accretion, $j/\alpha_{\rm v} c_s^2$,
is shortened by the radiation drag, where $j$ is the specific angular momentum,
$\alpha_{\rm v}$ is the viscous parameter, and $c_s$ is the sound velocity.
Thus, the MDO is likely to be a massive self-gravitating viscous disk.
For a self-gravitating viscous disk, some self-similar solutions
are known to give an inside-out disk collapse (Mineshige \& Umemura 1996;
Mineshige \& Umemura 1997; Tsuribe 1999).
In particular, Tsuribe (1999) provided a convenient formula for 
the inside-out mass accretion rate, 
\begin{equation}
\dot{M}_\alpha={3 \alpha_{\rm v} c_s^3 \over QG},
\end{equation}
where $c_s$ is the sound speed and 
$Q$ is the Toomre's Q which is $\kappa c_s/\pi G \Sigma$ for 
the epicycle frequency $\kappa$ and the surface density $\Sigma$.
Tsuribe (1999) has found a solution of stable accretion with
$Q \approx 2$. The critical accretion rate is given by
\begin{equation}
\dot{M}_\alpha \simeq {c_s^3 \over G}=0.24M_\odot {\rm yr}^{-1}
\left({ T \over 10^4 ~{\rm K}}\right)^{3/2}.
\end{equation}
This rate is again comparable to the Eddington mass accretion rate for
a black hole with $10^8 M_\odot$ if $T \approx 10^4$K
[see (\ref{m-dot-edd})].
Through this inside-out collapse, a central core grows. The core is
expected to be a rigidly rotating supermassive star because
the viscous transfer of angular momentum 
works to smear out any differential rotation in a self-gravitating system.

The equilibrium configuration and the stability of
a rigidly rotating supermassive star has been scrutinized by
Baumgarte \& Shapiro (1999). They found that a rotating supermassive star
becomes unstable for $R<640/GM/c^2$. 
As for the dynamical collapse of a rotating supermassive star,
Saijo et al. (2002) performed post-Newtonian calculations
and found that if a rotating supermassive star is in rigid rotation,
it can collapse towards the horizon scale without undergoing bar-mode
instability. The final stage of the collapse was investigated 
by Shibata \& Shapiro (2002)
with a full general relativistic approach and the emergence of
an apparent horizon of a Kerr black hole 
was shown. The resultant spin parameter of the black hole is
around 0.75.

Hence, an MDO formed by radiation 
drag-induced mass accretion could evolve into a supermassive Kerr black hole
through the inside-out viscous collapse and the general relativistic instability.
The present picture for the formation of a supermassive black hole
is summarized in Figure \ref{fig1}.

\subsection{Feedback by AGN Activity}

If the BH accretion causes the nuclear activity,
the further radiative mass accretion can be induced
by the nuclear luminosity $L_{\rm AGN}$. 
This feedback works positively to grow the SMBH.
Consequently, it is predicted that the BH mass to bulge mass ratio becomes
\begin{equation}
M_{\rm BH}/M_{\rm bulge}=0.14\varepsilon \alpha^{-1}(1-\eta)^{-1}
=0.003\alpha_{0.5}^{-1}. \label{m-ratio}
\end{equation}
This is just comparable to the observed ratio.

\subsection{$M-\sigma$ Relation}

If the BH mass is determined by the present mechanism, 
the BH mass to velocity dispersion relation is naturally understood
in the context of a cold dark matter (CDM) cosmology.
Supposing the bulge is a virialized system, 
then $GM_{\rm tot}/R = \sigma^2$ and
\begin{equation}
R \approx 0.5 R_{\rm max} \propto M_{\rm tot}^{1/3}(1+z_{\rm max})^{-1},
\end{equation}
where $R_{\rm max}$ is the radius at the maximum expansion epoch $z_{\rm max}$.
If a CDM cosmology is assumed, then 
\begin{equation}
(1+z_{\rm max}) \propto M_b^{-\beta}, 
\end{equation}
where $\beta \simeq 1/6$ around $M_b=10^{12}M_\odot$, 
almost regardless of the cosmological parameters (Bunn \& White 1997).
Combining all these relations, we find $M_{\rm BH} \propto \sigma ^n$
with $n=6/(2-3\beta)$, which is 4 for $\beta=1/6$.
This result is just corresponding to the inferred relation between 
the BH mass and the stellar velocity dispersion. 

\subsection{SMBH in Disks}

The radiation drag efficiency could be strongly subject to the 
effect of geometrical dilution (Umemura, Fukue, \& Mineshige 1998).
If the system is spherical, the emitted photons are effectively consumed 
within the system, whereas a large fraction of photons can escape 
from a disk-like system and also the drag efficiency is considerably
reduced across the optically-thick disk. 
This may be the reason why $f_{\rm BH}$ is
observed to be significantly smaller than 0.001 in disk systems.

\subsection{Hierarchy of Massive BHs}

In the present model, the theoretical upper limit of $f_{BH}$ 
is $\varepsilon=0.007$.
The low mass end of the BH hierarchy is suggested by the recent
observations on massive BHs in globular clusters, e.g., 
$M_{\rm BH} \approx 2 \times 10^4 M_\odot$ in G1 
(Gebhardt, Rich, \& Ho 2002) and
$M_{\rm BH} \approx 3 \times 10^3 M_\odot$ in M15 
(van der Marel et al. 2002; Gessen et al. 2002).
It is found that $M-\sigma$ relation still holds in such small 
spheroidal systems. 
The present radiation hydrodynamic mechanism
is expected to work in forming a BH even in a small spheroidal system, 
if the system retains interstellar medium (ISM) sufficient to become
optically thick. Although in globular clusters a few supernovae can
easily expel the ISM in an early evolutionary phase, 
the mass loss from low mass stars in later phases may accumulate a sufficient ISM.
On the other hand, M33 appears not to have a massive BH with an upper limit of 
$1500 M_\odot$ (Gebhardt et al. 2001).
An intriguing difference is that 
M33 contains a significant population of stars younger than a few
Gyrs (e.g., O'Connell 1983), in contrast to other globular clusters
which possess massive BHs. Thus, M33 might not have been able to
retain a sufficient ISM owing to successive supernova explosions.
In other words, the BH mass in globular clusters may 
sensitively reflect its star formation history.

As for the high mass end of the BH hierarchy, 
whether hypermassive BHs in clusters of galaxies exist
is a crucial test. 
In galaxy clusters, it is not likely that optically-thick intracluster
medium is supplied by galactic winds or stripping of ISM.
Hence, in the present scenario, hypermassive BHs 
will not form in the centers of galaxy clusters.
Therefore, the BH hierarchy is predicted to be cut off at
the maximum size in galaxies,
$M_{\rm BH} \approx 10^{10} M_\odot$.

\begin{figure}
\hspace{-1cm}
\includegraphics[width=12cm]{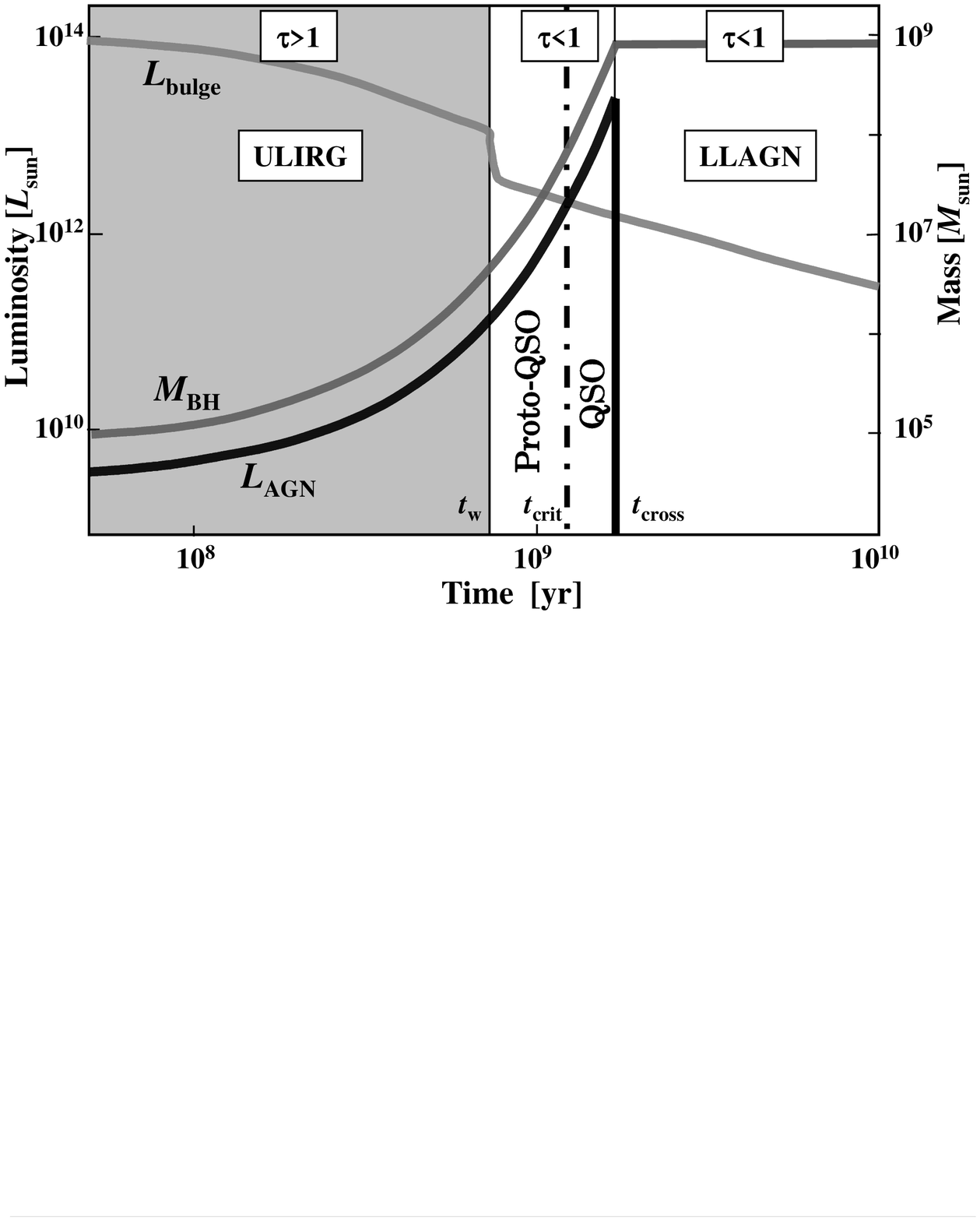}
    \caption{
A scenario for QSO formation. 
The abscissa is time and the ordinate is luminosity or mass.
$L_*$ and $L_{\rm AGN}$ are the bulge luminosity and
the black hole accretion luminosity, respectively.
$M_{\rm BH}$ is the mass of SMBH.
The ordinary QSO phase is preceded
by a host luminosity-dominant ``proto-QSO phase''.}
    \label{fig2}
\end{figure}

\section{QSO Formation}

\begin{figure}
    \centering
\hspace*{-1.5cm}
\includegraphics[width=13.2cm]{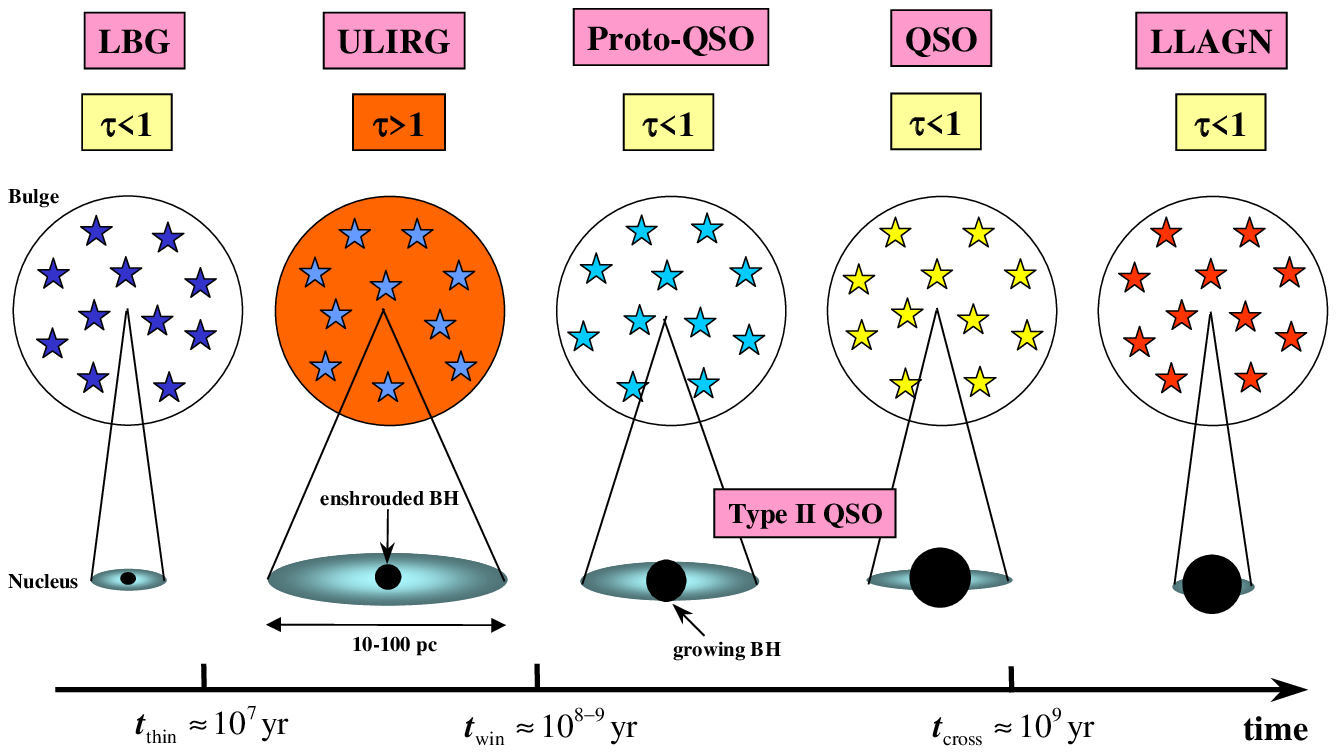}
    \caption{Schematic sketch for the coevolution of a SMBH and host galaxy. }
    \label{fig3}
\end{figure}

Here, we construct a picture of QSO formation based on the
present scenario for SMBH formation.
If the mass accretion driven by the viscosity on to the BH horizon 
is determined by an order of Eddington rate, 
the BH mass grows according to 
\begin{equation}
M_{\rm BH}=M_{0}e^{\nu t/t_{\rm Edd}},
\end{equation}
 where $\nu$ is the ratio of BH accretion rate to the Eddington rate
and $t_{\rm Edd}$ is the Eddington timescale, $t_{\rm Edd}=1.9\times 10^{8}$yr.  
$M_{0}$ is the mass of a seed BH with $\sim 10^{5}M_{\odot}$ 
(Shibata \& Shapiro 2002).
In order to incorporate the chemical evolution of host galaxy,
we use an evolutionary spectral synthesis code 'PEGASE' 
(Fioc \& Rocca-Volmerange 1997), and also employ a galactic wind model
with the wind epoch of $t_{\rm w}=7\times 10^8$yr to match the
present-day color-magnitude relation (Arimoto \& Yoshii 1987).
The system is assumed to change from optically-thick to optically-thin
phase at $t_{\rm w}$.
Based on the present coevolution model, 
the evolution of bulge luminosity ($L_{\rm bulge}$), 
the AGN luminosity ($L_{\rm AGN}$), and 
the mass of SMBH ($M_{\rm BH}$) 
are shown in Figure \ref{fig2}, 
assuming the constant Eddington ratio ($\nu =1$).
The $M_{\rm BH}$ reaches $M_{\rm MDO}$ at a time $t_{\rm cross}$,
so that the BH fraction becomes $f_{\rm BH}\simeq 0.001$, 
which is just comparable to the observed ratio.

In the optically-thin phase after the galactic wind ($t>t_{\rm w}$), 
$M_{\rm BH}$ still continues to grow until $t_{\rm cross}$ 
and therefore the AGN brightens with time.
After $L_{\rm AGN}$ exhibits a peak at $t_{\rm cross}$, 
it fades out abruptly due to exhausting the fuel of the MDO.
This fading nucleus could be a low luminosity AGN (LLAGN). 
It is found that the era of $t_{\rm w} < t <t_{\rm cross}$ 
can be divided into two phases with a transition time 
$t_{\rm crit}$ when $L_{\rm bulge}=L_{\rm AGN}$; 
the earlier phase is the host luminosity-dominant phase
and the later phase is the AGN luminosity-dominant phase.
The lifetimes of both phases are comparable to each other, 
which is about $10^{8}$yr.
The AGN-dominant phase is likely to correspond to ordinary QSOs, 
while the host-dominant phase is obviously different from observed QSOs so far.
We define this phase as ``a proto-QSO'' (Kawakatu \& Umemura 2003).
The observable properties of proto-QSOs are predicted as follows:
(1) The width of broad emission line is narrower, which is less than $1500$km/s.
(2) $f_{\rm BH}$ rapidly increases
from $10^{-5.3}$ to $10^{-3.9}$ in $\approx 10^{8}$ years.
(3) The colors of $({\it B-V})$ at rest bands and $({\it V-K})$ at observed bands 
are about 0.5 magnitude bluer than those of QSOs.
(4) In both proto-QSO and QSO phases, the metallicity of gas in galactic nuclei is 
$Z_{\rm BLR}\simeq 8Z_{\odot}$, 
and that of stars weighted by the host luminosity is $Z_{*}\simeq 3Z_{\odot}$,
which are consistent with the observations for QSOs and the elliptical galaxies.
(5) A massive dusty disk ($ > 10^{8}M_{\odot}$) surrounds a massive BH, 
and it may obscure the nucleus in the edge-on view to form a type 2 nucleus.
The predicted properties of proto-QSOs are quite similar to those of 
radio galaxies at high redshifts. 
Thus, high-$z$ radio galaxies are a key candidate for proto-QSOs.

The proto-QSO phase is preceded by a bright and optically thick phase,
which may correspond to a ultraluminous infrared galaxy (ULIRG) phase.
Also, the precursor of ULIRGs is an optically-thin and very luminous
phase with the lifetime of $\sim 10^{7}$ years. 
This may correspond to the assembly phase of LBGs or Ly$\alpha$ emitters.
In this phase, the metallicity is subsolar ($Z_{*} < 0.1Z_{\odot} $), 
and the hard X-ray luminosity is $L_{\rm x}\sim 5\times 10^{8}L_{\odot}$ 
if $L_{\rm x}=0.1L_{\rm AGN}$.
Such a coevolution scenario of a SMBH and the host is summarized
in Figure \ref{fig3}.

\section{Conclusions }

The SMBH could form through the relativistic radiation-hydrodynamic processes
from galactic scale to horizon scale.
The resultant mass of a SMBH is in proportion
to the bulge mass, and the SMBH mass fraction is determined by 
the nuclear energy conversion efficiency from hydrogen to helium, 
$\varepsilon=0.007$, which is the theoretical upper limit in the present
mechanism. Massive BHs can form also in globular clusters, but 
hypermassive BHs are unlikely to form in the centers of galaxy clusters.
As for QSO formation, it is predicted that 
a host luminosity-dominant ``proto-QSO phase'' exists before the QSO phase.
The proto-QSO is a phase of growing BH with broad emission line width of
less than $1500$km/s.
The proto-QSO phase is preceded by an optically-thick ultraluminous infrared
galaxy (ULIRG) phase. Before the ULIRG phase, an optically thin 
starburst phase should exist.

\end{document}